\def\tn{\textnumero}
\def\BE{\begin{equation}}
\def\EE#1{\label{#1}\end{equation}}
\def\be{\begin{align}}
\def\ba{\begin{align*}}
\def\se#1{\begin{subequations}\label{#1}
\renewcommand{\theequation}{\theparentequation\alph{equation}}}
\def\rf#1{(\ref{#1})}
\def\I{{\rm i}}
\def\e{{\rm e}}
\def\D{{\rm d}}

\def\R{{\mathbb R}^3}
\def\T{{\mathbb T}^3}
\def\om{\boldsymbol{\omega}}
\def\P{\mathfrak{P}_{\bf n}}
\documentclass[12pt]{Melsarticle}
\usepackage{mathtools,amsmath,amssymb,textcomp,mathrsfs}
\begin{document}
\title{Depletion of nonlinearity in space-analytic space-periodic solutions
to equations of diffusive magnetohydrodynamics}
\author[mitpan]{V.~Zheligovsky}
\address[mitpan]{Institute of Earthquake Prediction Theory and
Mathematical Geophysics, Russian Ac. Sci.,\\
84/32 Profsoyuznaya St, 117997 Moscow, Russian Federation}

\begin{abstract}
We consider solenoidal space-periodic space-analytic solutions to the equations
of magnetohydrodynamics. An elementary bound shows that due to the special
structure of the nonlinear terms in the equations for modified solutions,
effectively they lack a half of the spatial gradient, which appears to be
a novel mechanism for depletion of nonlinearity. We
present a two-phase iterative procedure yielding an expanded bound
for the guaranteed time of the space analyticity of the hydrodynamic solutions.
Each iteration involves two regimes: In phase 1, the enstrophy of the modified
solution and the bound for the radius of the analyticity of the original
solution simultaneously increase (the bound is proportional to the elapsed time
since the beginning of phase~1). In phase 2, the enstrophy and bound
simultaneously decrease. It is straightforward to generalize this construction
for the equations of magnetohydrodynamics.
\end{abstract}
\maketitle

\section{Introduction}
The simultaneous evolution of the velocity, ${\bf V(x},t)$, of an electrically
conducting fluid flow and the magnetic field, ${\bf B(x},t)$, satisfies
the Navier--Stokes equation involving the Lorentz force,
\se{vb}\be\partial{\bf V}/\partial t&=\nu\nabla^2{\bf V}
+{\bf V}\times\om-{\bf B}\times(\nabla\times{\bf B})-\nabla P,\quad\label{ve}\\
\shortintertext{and the magnetic induction equation,}
\partial{\bf B}/\partial t&=\eta\nabla^2{\bf B}
+\nabla\times({\bf V}\times{\bf B}).\label{be}\\
\shortintertext{The fluid is supposed to be incompressible and
the magnetic field is solenoidal,}
\nabla\cdot{\bf V}&=\nabla\cdot{\bf B}=0.\label{bvs}\end{align}\end{subequations}
Here $P$ denotes the modified pressure, $\om=\nabla\times\bf V$ the vorticity,
${\bf x}\in\R$ are the Eulerian coordinates and $t$ is time. For the sake
of simplicity, no external forcing is assumed. We consider solutions
in $\T=[0,2\pi]^3$, which are $2\pi$-periodic in each Cartesian variable $x_i$.

The nonlinear terms in the equations \rf{vb} involve numerous vector products
(including the curl operators, which also reduce to vector products in the
Fourier space) resulting in significant cancellations and having implications
for the structures developing in the two physical fields. This phenomenon,
called the depletion of nonlinearity (see \cite{Fr}), was studied extensively
in the simulations of turbulence in two (e.g., see \cite{Pu}) and three dimensions.
In hydrodynamics, it can be associated with the flow beltramization
in the areas of large vorticity, i.e., the velocity and vorticity tend
to approximately line up, whereby their vector product can be significantly
smaller than on the periphery. In these regions, the depletion
can also be caused by the flow self-organization into the structures of reduced
dimensionality such as
one-dimensional ropes \cite{Fr}. Furthermore, for a random solenoidal
Gaussian vector field $\bf V$ modelling turbulence, the potential part
of the Lamb vector ${\bf V}\times\om$ can exceed twice the solenoidal one
(in the sense of the mean-square norm) \cite{Tsi}. In solutions
to the full magnetohydrodynamic (MHD) problems, the flow and magnetic field
show a tendency to become
parallel in the areas of large gradients \cite{Ma,Se}, and $\bf V\approx B$
in the extreme case of the Archontis dynamo~\cite{DA,CG,Ga}.

Also, this cancellation has mathematical consequences. The analysis
\cite{Do,G14} (see also \cite{Gi}) of solutions
to the Navier--Stokes equation computed with a high spatial resolution (up to
$8192^3$ Fourier harmonics) revealed an unexpected monotonic growth
in the quantities $D_m$ (that are powers of the scaled moments of the vorticity)
on increasing $m$;
this was interpreted as a manifestation of the nonlinearity weakening.
A higher depletion was found in numerical space-periodic simulations
of magnetohydrodynamic (MHD) turbulence \cite{G16}, where similar quantities
$D^\pm_m$ computed for the curls of the Els\"asser variables $\bf V\pm B$
were studied (see also \cite{Je}).

The finite-time spatial analyticity of space-periodic solutions
to the Navier--Stokes equation was demonstrated in \cite{FT} by deriving bounds
for their Gevrey class norms (see \cite{BGK} for a review of results
on the analyticity of solutions
to the hydrodynamic problem). Moreover, if a three-dimensional space-periodic
solution to the MHD equations \rf{vb} belongs initially to the Sobolev space
$H_{1/2}(\T)$, it instantly acquires space analyticity \cite{Zh} (see also
\cite{Ki}) and subsequently it is almost always space-analytic
\cite{Zh}.~The proofs \cite{Zh} relied on using the modified solutions
\se{mo}\be{\bf v}&=\sum_{{\bf n}\ne 0}{\bf v_n}(t)\e^{\I\bf n\cdot x},\quad
{\bf V_n=v_n}\e^{-\Gamma|{\bf n}|},\label{mv}\\
{\bf b}&=\sum_{{\bf n}\ne 0}{\bf b_n}(t)\e^{\I\bf n\cdot x},\quad
{\bf B_n=b_n}\e^{-\Gamma|{\bf n}|},\label{mb}\end{align}\end{subequations}
where $\bf V_n$ and $\bf B_n$ are the Fourier coefficients of $\bf V$ and
$\bf B$, respectively, and $\Gamma\ge0$ is independent of the three-dimensional
wave vector $\bf n$ (see also \cite{LO,Zh11}).
If the modified solution is smooth, $\Gamma$ is a lower bound
for the radius of the space analyticity of the original solution.
The time integrals of certain powers of high-index Sobolev norms of hydrodynamic
solutions were shown in \cite{FGT} to be finite.~This was generalized \cite{Zh}
to encompass MHD solutions for space-analytic initial conditions
by employing
$$\Gamma=\delta(1+\|{\bf v}\|_{3/2}^2+\|{\bf b}\|_{3/2}^2)^{-1/2}$$
in \rf{mo}. Here $\|\cdot\|_p$ denotes the norm in the Sobolev
space $H_p(\T)$ and $\delta>0$ is a constant.

The identities
$${\bf c}\times(\nabla\times{\bf c})=-({\bf c}\cdot\nabla){\bf c}
+{1\over2}\nabla|{\bf c}|^2,\qquad
\nabla\times({\bf c}\times{\bf d})=({\bf d}\cdot\nabla){\bf c}
-({\bf c}\cdot\nabla){\bf d},$$
which hold true for any solenoidal fields $\bf c$ and $\bf d$, attest that
all the nonlinear terms in \rf{vb} can be expressed as the linear combinations
of directional derivatives (of the type $(\bf d\cdot\nabla)c)$. The Fourier
coefficients $\bf v_n$ and $\bf b_n$ are governed by the ordinary differential
equations
\se{mvb}\be{\D{\bf v_n}\over\D t}&+\nu|{\bf n}|^2{\bf v_n}
-|{\bf n}|{\bf v_n}\,{\D\Gamma\over\D t}
=\I\sum_{\bf k}\e^{\Gamma(|{\bf n}|-|{\bf k}|-|{\bf n-k}|)}
\P\big(({\bf b_k}\cdot{\bf n}){\bf b}_{\bf n-k}
-({\bf v_k}\cdot{\bf n}){\bf v}_{\bf n-k}\big),\label{vo}\\
{\D{\bf b_n}\over\D t}&+\eta|{\bf n}|^2{\bf b_n}
-|{\bf n}|{\bf b_n}\,{\D\Gamma\over\D t}
=\I\sum_{\bf k}\e^{\Gamma(|{\bf n}|-|{\bf k}|-|{\bf n-k}|)}
\big(({\bf b_k}\cdot{\bf n}){\bf v}_{\bf n-k}
-({\bf v_k}\cdot{\bf n}){\bf b}_{\bf n-k}\big),\label{bo}\end{align}
\end{subequations}
where $\P$ is the projection of a three-dimensional vector on the plane,
orthogonal to $\bf n$.

We present here an alternative mechanism of the depletion of nonlinearity.
In the next section we prove an elementary estimate, implying that when
majorizing the nonlinear terms in \rf{mvb}, the factor $\bf|n|$ can be replaced
by $(|{\bf n-k}|^{1/2}+|{\bf n-k}||{\bf k}|^{-1/2})\Gamma^{-1/2}$, i.e.,
although formally the gradient is linear in the wave number, effectively a half
of it disappears in the bound. Thus, it is shown that the nonlinearity depletes,
inter alia, due to the reduction in the effective order of the differential
operators involved. By contrast, the factor $\Gamma^{-1/2}$ increases
the advective terms, when the bound $\Gamma$ is small. To the best of our
knowledge, this
mechanism of the nonlinearity depletion was never considered in the literature
previously. We apply it to estimate the interval of the guaranteed existence
of the space-analytic solution in section \ref{ge}.

\section{A bound for nonlinear terms}

We demonstrate here the inequality
\BE|{\bf(c_k\cdot n)\P d_{n-k}}|\e^{\Gamma({\bf|n|-|n-k|-|k|})}
\le{|\bf c_k||d_{n-k}|}\sqrt{{\bf|n-k|}\over\Gamma}
\left(1+\sqrt{\bf|n-k|\over|k|}\right)\EE{trm}
for the nonlinear terms in the sums in the r.h.s.~of Eqs.~\rf{mvb};
$\bf c$ and $\bf d$ are ``metavariables'' standing for $\bf v_m$ and/or
$\bf b_m$ (the projection may be unused in the case of \rf{bo}).

Since $\bf c$ is solenoidal,
${\bf c_k=|k}|^{-2}\bf k\times(c_k\times k)$ implying
$$|{\bf c_k\cdot(n-k)}|\e^{\Gamma(\bf|n|-|n-k|-|k|)}\le|\varphi(\theta)|
{\bf|c_k||n-k|},$$
where
$$\varphi(\theta)=\sin\theta\,\exp(\Gamma({\bf|n|-|n-k|-|k|}))$$
and $\theta$ is the angle between the vectors $\bf n-k$ and $\bf k$.
At the point $\theta=\theta_{\max}$ of the maximum of $|\varphi(\theta)|$,
the necessary condition for an extremum, $\D\varphi/\D\theta=0$, holds true,
which is equivalent to the relation
\be\sin^2\theta_{\max}\,\Gamma{\bf|n-k||k|}&=-\cos\theta_{\max}{\bf|n|}
\left.\vphantom{|_|}\right|_{\theta=\theta_{\max}}\nonumber\\[-.5ex]
&\le{\bf|n-k|+|k|}.\label{si}\end{align}
By the triangle inequality, the exponent in the definition of $\varphi(\theta)$
is non-positive and hence the exponential is majorized by the unity; now,
by \rf{si},
\BE|\varphi(\theta)|\le|\varphi(\theta_{\max})|\le|\sin\theta_{\max}|\le
\bigg({{\bf|n-k|+|k|}\over\Gamma{\bf|n-k||k|}}\bigg)^{1/2},\EE{in}
which proves \rf{trm}.

{\it Remark.}
Similar arguments prove the inequality
$$|{\bf(c_k\cdot n)\P d_{n-k}}|\e^{\Gamma({\bf|n|-|n-k|-|k|})}\le{|\bf c_k||d_{n-k}|}
\big({\bf|n|}^2+{\bf|n|}^4/{\bf|k|}^2\big)^{1/4}\,\Gamma^{-1/2}.$$

\section{Interval of the guaranteed existence of a space-analytic solution}
\label{ge}

For simplicity, we consider in this section exclusively the hydrodynamic
problem for ${\bf B}=0$. Since all nonlinear terms in \rf{mvb} have
the same structure, it is easy to obtain similar bounds for the MHD problem
following the same approach. We assume \hbox{$\|{\bf V}\|_1<\infty$} at $t=0$.

A variant of the bound implying the flow space analyticity upon its
spontaneous emergence \cite{FT} can be obtained for $\Gamma=\nu^{-1}t$.
Applying the embedding theorem, we deduce from \rf{vo} for ${\bf B}=0$
$${1\over2}{\D\over\D t}\|{\bf v}\|_1^2
+\nu\|{\bf v}\|_2^2-\nu^{-1}\|{\bf v}\|_{3/2}^2
\le\widetilde{C}\|{\bf v}\|_2\|{\bf v}\|_{3/2}\|{\bf v}\|_1,$$
and hence, by the H\"older and Young inequalities,
\BE{\D y^2\over\D t}\le C\nu^{-3}y^2(1+y^2)\quad\Rightarrow\quad y^2\le
\bigg(\!\bigg(1+{1\over y^2(0)}\bigg)\e^{-C\nu^{-3}t}-1\bigg)^{-1},\EE{y}
where $\widetilde{C}$ and $C$ are the relevant constants and
$y=\|{\bf v}\|_1^2$. This inequality guarantees the existence
of the space-analytic solution until
$$T_*=\nu^3C^{-1}\ln(1+1/y^2(0)).$$

Let us consider a process, which is in some sense reciprocal. Scalar multiplying
\rf{vo} by ${\bf v_{-n}|n}|^2$, summing the results over $\bf n$, transforming
identically the term involving $\D\Gamma/\D t$ and applying \rf{trm}
to the sum in the r.h.s., we obtain
\be&{1\over2}{\D\over\D t}\|{\bf v}\|_1^2+\nu\|{\bf v}\|_2^2\label{go}\\
\le&\|{\bf v}\|_{3/2}^2\,\Gamma^{-1/2}\Biggl(\!{2\over3}{\D\Gamma^{3/2}\over\D t}
+\|{\bf v}\|_{3/2}^{-2}\sum_{\bf n,k}
|{\bf v}_{\bf k}||{\bf v}_{\bf n-k}||{\bf v}_{\bf-n}|{\bf|n|}^2
\left({\bf|n-k|}^{1/2}+{{\bf|n-k|}\over|{\bf k}|^{1/2}}\right)\!\Biggr).
\nonumber\end{align}
The sum, $\Sigma$, in the r.h.s.~can be estimated
applying the embedding theorem and the H\"older inequality as follows:
\ba\Sigma
&\le\sum_{\bf n,k}|{\bf v}_{\bf k}||{\bf v}_{\bf n-k}||{\bf v}_{\bf-n}|
{\bf|n|}^{3/2}\Big(2{\bf|n-k|}+{\bf|n-k|}^{1/2}{\bf|k|}^{1/2}
+{\bf|n-k|}^{3/2}{\bf|k|}^{-1/2}\Big)\\
&=(2\pi)^{-3}\int_{\T}\big(2u_{3/2}u_1u_0+u_{3/2}u_{1/2}^2+u_{3/2}^2u_{-1/2}
\big)\D{\bf x}\\
&\le(2\pi)^{-3}\big(2|u_{3/2}|_2|u_1|_3|u_0|_6+|u_{3/2}|_2|u_{1/2}|_3|u_{1/2}|_6
+|u_{3/2}|_2|u_{3/2}|_{6/(3-\alpha)}|u_{-1/2}|_{6/\alpha}\big)\\
&\le C'\|{\bf v}\|_{3/2}^2\|{\bf v}\|_1
+C'_\alpha\|{\bf v}\|_{3/2}\|{\bf v}\|_{(3+\alpha)/2}\|{\bf v}\|_{1-\alpha/2}\\
&\le C_\alpha\|{\bf v}\|_2^\alpha\|{\bf v}\|_{3/2}^{2-\alpha}\|{\bf v}\|_1\end{align*}
\vspace*{-2em}
\BE\Rightarrow\quad\|{\bf v}\|_{3/2}^{-2}\Sigma\le C_\alpha\|{\bf v}\|_2^\alpha
\|{\bf v}\|_1^{1-\alpha}.\EE{gi}
Here
$$u_p=\sum_{\bf m}|{\bf v_m}||{\bf m}|^p\e^{\,\I\bf m\cdot x},$$
$|\cdot|_q$ is the norm in the Lebesgue space $L_q(\T)$, $C',C'_\alpha$ and
$C_\alpha$ are the relevant constants, $\alpha$~is confined to the interval
$1>\alpha>0$ (we note $C_\alpha\to\infty$ when $\alpha\to0$).

By \rf{gi}, the norms in the l.h.s.~of \rf{go} are controlling the sum
in the r.h.s.~of \rf{go}. This suggests an iterative procedure for bounding
the modified solution. Each iteration consists of two phases. We use
the following notation to describe the procedure:
the start time of the $j$th iteration is denoted by $T_j$,
the durations of the two phases comprising it by $\lambda^{(1)}_j$ and
$\lambda^{(2)}_j$; the second phase begins at $T'_j=T_j+\lambda^{(1)}_j$;
hence, $T_{j+1}=T'_j+\lambda^{(2)}_j$. Finally, we denote by $\Lambda^{(m)}_j$
the lower bounds that we derive for $\lambda^{(m)}_j$.

{\it Phase 1. The simultaneous growth of the enstrophy of the modified solution
and of the bound $\Gamma$ for the radius of analyticity of the original
solution.} At $t=T_j$ we switch on
the regime of the spontaneous emergence of the analyticity until the enstrophy
$\|{\bf v}\|_1^2$ increases at time $T'_j=T_j+\lambda^{(1)}_j$ by a factor $a_j>1$.
By~\rf{y}, the duration of this phase satisfies the inequality
\BE\lambda^{(1)}_j\ge\Lambda^{(1)}_j=\nu^3C^{-1}\ln\big(a_j^2(1+y_j^2)/(1+a_j^2y_j^2)\big),\EE{d}
where $y_j=\|{\bf v}(T_j)\|_1^2$.~Consequently,
$\Gamma\ge\nu^{-1}\Lambda^{(1)}_j$ at $t=T'_j$.

{\it Phase 2. Simultaneous decrease in the enstrophy and bound $\Gamma$.}
At $t=T'_j$ we switch on the reciprocal regime of the bound decrease
according to the equation
\BE-{2\over3}{\D\Gamma^{3/2}\over\D t}=
C_\alpha\|{\bf v}\|_2^\alpha\|{\bf v}\|_1^{1-\alpha}\EE{Gf}
until $\Gamma=0$ at time $T_{j+1}=T'_j+\lambda^{(2)}_j$. By \rf{go} and \rf{gi},
$${1\over2}{\D\over\D t}\|{\bf v}\|_1^2+\nu\|{\bf v}\|_2^2\le0,$$
implying
\BE\int_{T'_j}^t\|{\bf v}\|_2^2\,\D t\le a_jy_j/\nu;\qquad
\|{\bf v}\|_1^2\le a_jy_j\e^{-2\nu(t-T'_j)}\EE{bb}
(the second inequality holds true because $\|{\bf v}\|_1\le\|{\bf v}\|_2$).

To estimate $\lambda^{(2)}_j$, we integrate~\rf{Gf} from $T'_j$ to $T_{j+1}$
exploiting the relations $\Gamma(T_{j+1})=0$ ($T_{j+1}=\infty$, if this
value is unreachable) and \rf{bb}, and the H\"older inequality:
\ba{2\over3}\big(\nu^{-1}\Lambda^{(1)}_j\big)^{3/2}&\le{2\over3}\,\Gamma^{3/2}(T'_j)
=\int_{T'_j}^{T_{j+1}}C_\alpha\|{\bf v}\|_2^\alpha\|{\bf v}\|_1^{1-\alpha}
\D t\\[-1ex]
\le&\,C_\alpha\left(\int_{T'_j}^{T_{j+1}}\|{\bf v}\|_2^2\,
\D t\right)^{\alpha/2}\left(\int_{T'_j}^{T_{j+1}}\|{\bf v}\|_1^\mu\D t
\right)^{1-\alpha/2}\\[-1ex]
\le&\,C_\alpha(a_jy_j)^{1/2}\nu^{-\alpha/2}\left(\int_{T'_j}^{T_{j+1}}
\e^{-\nu\mu t}\D t\right)^{1-\alpha/2}\\[-.5ex]
=&\,C_\alpha(a_jy_j)^{1/2}\nu^{-1}\mu^{\alpha/2-1}\big(1-
\e^{-\nu\mu\lambda^{(2)}_j}\big)^{1-\alpha/2},\end{align*}
where $\mu=(1-\alpha)/(1-\alpha/2)$. Thus, the duration of phase~2
satisfies the inequality
\se{dp}\be\lambda^{(2)}_j&\ge\Lambda^{(2)}_j=-(\nu\mu)^{-1}\ln Q_j,\label{di}\\[-1ex]
\shortintertext{where}
Q_j&=1-\mu\bigg({4\nu^8\ln^3\big(a_j^2(1+y_j^2)/
(1+a_j^2y_j^2)\big)\over9C^3C^2_\alpha a_jy_j}\bigg)^{1/(2-\alpha)};
\label{qq}\end{align}\end{subequations}
$\lambda^{(2)}_j=T_{j+1}=\infty$, if $Q_j\le0$.

By \rf{bb}, we can now set
\BE y_{j+1}=a_j\e^{-2\nu\Lambda^{(2)}_j}y_j.\EE{jp}

\section{Conclusions}

We have presented a new mechanism of the nonlinearity depletion (weakening)
for space-analytic
solutions to the MHD equations. It is manifested by the inequality \rf{trm},
whereby the directional derivative $\bf(c\cdot\nabla)d$ for solenoidal vector
fields $\bf c$ is bounded by a homogeneous function of wave vector
lengths of degree 1/2. The inequality involves the lower bound $\Gamma$
for the radius of the analyticity
of the solution. A~qualified reduction of $\Gamma$ has been shown to cause
a decrease in the enstrophy of the modified solution. We have designed
a two-phase iterative procedure for estimating the guaranteed time
of the existence
of space-analytic solutions that exploits the two mechanisms. It involves
two alternating regimes: a simultaneous increase in phase 1 in the enstrophy
of the modified solution with a linear in time growth of the bound $\Gamma$
for the radius of the analyticity of the original solution (the regime
discovered in \cite{FT}), and a simultaneous decrease in the enstrophy
and $\Gamma$ in phase~2.

We have established that when the initial flow enstrophy is finite,
a solution to the Navier--Stokes equation remains space-analytic at least up to
$T_{**}=\sum_j(\Lambda^{(1)}_j+\Lambda^{(2)}_j)$, see \rf{d} and \rf{dp}.
Formally, in our construction $\Gamma$ is allowed to decrease to zero when
phase~2 terminates; however, the enstrophy of the flow (which does not exceed
the enstrophy of the modified solution) is uniformly bounded on the intervals
$[0,T_{**}-\varepsilon]$ for any \hbox{$\varepsilon>0$} and this guarantees the space
analyticity of the flow on any open interval $(0,T_{**})$. This expands
the guaranteed life-span $T_*$ \cite{FT} of the analytical solution.
In particular, \hbox{$T_*<\sum_j\Lambda^{(1)}_j$:} while $T_*$ is the sum of times,
during which a solution to the ODE
$$\D y^2/\D t=C\nu^{-3}y^2(1+y^2)$$
(cf.~\rf{y}) grows monotonically from $y_j$ to $y_{j+1}$, each
$\Lambda^{(1)}_j$ is the time it takes it to grow from
$y_j$ to a larger value $\e^{2\nu\lambda^{(2)}_j}y_{j+1}$. For small~$\nu$,
$\Lambda^{(1)}_j={\rm O}(\nu^3)$ and $\Lambda^{(2)}_j={\rm O}(\nu^{8/(2-\alpha)-1})$,
i.e., in terms of the orders of $\nu$ the newly introduced
phases 2 contribute almost as much to the length of the guaranteed interval
of the space analyticity as phases 1 (since any value
of the parameter $\alpha$ from the interval $1>\alpha>0$ is permitted).

By \rf{jp}, the enstrophy of the modified solution \rf{mv} changes during
the $j$th iteration by a factor equal to or below $a_j\e^{-2\nu\Lambda^{(2)}_j}$.
This product does not exceed the unity (enabling the iterations to have fixed
or increasing durations, thus implying the existence of a global analytical
solution), if and only if the initial enstrophy $y_j=\|{\bf v}(T_j)\|_1^2$
is sufficiently small or the viscosity $\nu$ is sufficiently large.

Some intriguing questions remain open: What is the optimal choice of the factors
$a_j$ maximizing the guaranteed life-span of the solution $T_{**}$?
Is letting $\Gamma$ decay in phase~2 according to \rf{bb} an optimal
strategy? If $\Gamma$ decreases faster, then phase 2 is
shorter, resulting in the larger factor $\e^{-2\nu\Lambda^{(2)}_j}$, but
the decrease in the enstrophy is enhanced by the negative r.h.s.~of \rf{go},
involving two factors: $\|{\bf v}\|_{3/2}^2$, which can be large and
$\Gamma^{-1/2}$, which is large towards the ends of phases~2.
So, what is the optimal control of $\Gamma$ in phase 2?
Furthermore, is there a better strategy than implementing the alternating
$\Gamma$ growth from 0 and decay to 0 as we have assumed here?

For the full system of the equations of the diffusive magnetohydrodynamics,
the derivations are similar, since all the nonlinear terms have the same
structure of the directional derivatives along solenoidal fields.

\section*{Acknowledgements}
The work was financed by the grant \tn 22-17-00114
of the Russian Science Foundation, https:/\!/rscf.ru/project/22-17-00114/~.

\end{document}